\documentstyle[12pt]{article}

\parskip=\medskipamount                 
\begin{document}
\bibliographystyle{unsrt}
\begin{center}
{\large \bf Scaling limit of a non-relativistic model\\
of confined ``quarks''}\\
[.1in]
{\bf O.W. Greenberg\footnote{Supported in
part by the National Science Foundation.\\
e-mail addresses: greenberg@umdhep (bitnet)\\
umdhep::greenberg (decnet)\\
greenberg@umdep1.umd.edu (internet)}}\\[.1in]
{\it Center for Theoretical Physics}\\[-.03in]
{\it Department of Physics and Astronomy}\\[-.03in]
{\it University of Maryland}\\[-.03in]
{\it College Park, MD~~20742-4111}\\
Preprint number 92-223\\
[.6in]
\end{center}
{\bf Abstract}\vglue.1in
I calculate the structure function for scattering from the two-body bound
state in its lowest level in a non-relativistic model of confined scalar
``quarks'' of masses $m_A$ and $m_B$.  The scaling limit in
$x={\bf q}^2/2(m_A+m_B)q^0$
exists and is non-vanishing only for the values $x=m_A/(m_A+m_B)$ and
$x=m_B/(m_A+m_B)$ which correspond to the fractions of the momentum of the
two-body system carried by each of the ``quarks.''  In the scaling limit,
the interference from scattering off of the two ``quarks'' vanishes.
Thus the scaling limit of this model agrees with the parton picture.
\newpage
\vspace{.6in}
\begin{center}
{\bf 1. INTRODUCTION}
\end{center}

Since the scaling limit of the structure functions for deep
inelastic scattering concerns a regime which is at an opposite extreme from
the confinement regime, it is interesting to see how asymptotic freedom,
which is responsible for
the high momentum transfer
scaling limit of deep inelastic scattering given by the parton model
\cite{ioffe},
coexists with confinement, which is of crucial importance for low-energy
hadronic physics.  In the present paper, I
study the scaling regime in a non-relativistic model of
scalar ``quarks'' bound by an harmonic potential to get some clues as to how
these disparate regimes can coexist in such models.  Insight gained might be
useful in studying the corresponding problem in QCD.
Specifically, I calculate the structure function
$W_{00}^{(AB)}$ for inelastic lepton scattering from a bound state of
``quarks''  of masses $m_A$ and $m_B$.
I find that the scaling limit in $x={\bf q}^2/2(m_A+m_B)q^0$
exists and is non-vanishing only for the values $x=m_A/(m_A+m_B)$ and
$x=m_B/(m_A+m_B)$ which correspond to the fractions of the momentum of the
two-body system carried by each of the ``quarks.''
For these values of $x$ the
scattering is that expected from free ``quarks.''  Thus the scaling limit of
this model agrees with the parton picture.
Because particles interacting via
an harmonic potential are much freer at short distances
than those which interact with a potential which is linear at large distances
and Coulombic with a
coupling constant which decreases logarithmically, as in the case of
asymptotic freedom, at short distances, one can expect a rapid approach to the
scaling limit.

Section 2 describes the model.  Section 3 contains the
calculation of the structure function.
Section 4 summarizes the work and gives the outlook for future developments.

With a view to possible later work, I carried out the calculation using the
N quantum approximation (NQA), which is exact in this model.  A brief review
of the NQA in the form relevant to theories with a confining potential
is given in the Appendix.
I emphasize that the NQA is not necessary for this calculation; the usual
Schr\"odinger theory gives the same result.

\begin{center}
{\bf 2. DESCRIPTION OF THE MODEL}
\end{center}

The model has two species of nonrelativistic scalar ``quarks'' of masses
$m_A$ and
$m_B$. (I could have called one of these a ``quark'' and the other an
``antiquark.'') The second-quantized
Hamiltonian of the model is
\begin{eqnarray}
H&=&\sum_{i=A,B}(2m_i)^{-1}\int d^3x \nabla\psi_i^{\dagger}({\bf x},t)\cdot
\nabla\psi_i({\bf x},t)  \nonumber \\
 & &+ (k/2)\int d^3xd^3y\psi_A^{\dagger}({\bf x},t)\psi_B^{\dagger}({\bf y},t)
({\bf x}-{\bf y})^2
\psi_B({\bf y},t)\psi_A({\bf x},t).      \label{1}
\end{eqnarray}
The Heisenberg equation of motion for $\psi_A$ is
\begin{equation}
i\partial\psi_A({\bf x},t)/\partial t=-(2m_A)^{-1}\nabla^2\psi_A({\bf x},t)
+k\int d^3y\psi_B^{\dagger}({\bf y},t)({\bf x}-
{\bf y})^2\psi_B({\bf y},t)\psi_A({\bf x},t),   \label{2}
\end{equation}
with a similar equation for $\psi_B$.  The solution of Eq.(\ref{2}) using the
Haag expansion is given in the Appendix.
The equation for the two-body Schr\"odinger
amplitudes, which are identical to the Haag amplitudes described in the
Appendix, in terms of the relative coordinate ${\bf r}={\bf x}-{\bf y}$, is
\begin{equation}
(-\frac{\hbar^2 \nabla^2_r}{2\mu_{AB}}+k{\bf r}^2)F_{AB,{\bf n}}({\bf
r})=\epsilon_{{\bf n}}F_{AB,{\bf n}}({\bf r}),  \label{4}
\end{equation}
\begin{equation}
\frac{1}{\mu_{AB}}=\frac{1}{m_A}+\frac{1}{m_B}.          \label{5}
\end{equation}
To fix notation, the solution of Eq. (\ref{4}) is $F_{AB,{\bf n}}({\bf
x})
=\prod_1^3F_{AB,{n_i}}(x_i)$,
\begin{equation}
F_{AB, n}(x)=\frac{1}{\sqrt{2^n n!}}(\frac{\mu_{AB}\omega}{\pi \hbar})^{1/4}
exp(-\frac{\mu_{AB} \omega x^2}{2\hbar})
H_n(\sqrt{\frac{\mu_{AB} \omega}{\hbar} x}),   \label{6}
\end{equation}
where $H_n$ is the Hermite polynomial.

\begin{center}
{\bf 3. CALCULATION OF THE STRUCTURE FUNCTION}
\end{center}

A general formula for the structure function is
\begin{eqnarray}
W_{\mu \nu}&=&\sum_{\alpha}\int d^3k
\langle {\bf P},{\bf 0}|j_{\mu}({\bf 0})|{\bf k},{\alpha}\rangle
\langle {\bf k},{\alpha} |j_{\nu}({\bf 0})|{\bf P},{\bf 0}\rangle (2 \pi)^4
\delta^4(q+P-P_{{\bf k},\alpha})  \nonumber \\
&=&\sum_{\alpha}\langle {\bf P},{\bf 0}|j_{\mu}({\bf 0})|{\bf P}+{\bf q},\alpha
\rangle \langle {\bf P}+{\bf q}, \alpha|j_{\nu}({\bf 0})|{\bf P},{\bf 0}\rangle
                                     \nonumber \\
           & &\times(2 \pi)^4
\delta(q^0+E_{0}({\bf P})-E_{\alpha}({\bf P}+{\bf q})) \label{7},
\end{eqnarray}
$\alpha$ labels the quantum numbers of the intermediate states aside from
the momentum.  For the non-relativistic model, I calculate only $W_{00}$.  I
work in the rest frame of the target, ${\bf P}={\bf 0}$; however, I will
put ${\bf P}$ in the label of the states, and set
${\bf P}={\bf 0}$ in the $\delta$-functions and in the final results, because
putting ${\bf 0}$ in the labels of the states might confuse the states with the
vacuum.
The requirement that the
charges $Q_i=\int d^3x \rho_i(x)$ obey
$\langle {\bf p}, {\bf n}|Q_i|{\bf p}^{\prime}, {\bf n}^{\prime}\rangle
=\delta_{{\bf n},{\bf n}^{\prime}}\delta({\bf p}-{\bf p}^{\prime})$
provides the normalization of the bound-state amplitudes,
\begin{equation}
\int d^3r F^{\ast}_{AB,{\bf n}}({\bf r})F_{AB,{\bf n}^{\prime}}
({\bf r})=\delta_{{\bf n}, {\bf n}^{\prime}}.  \label{12}
\end{equation}
The structure function for
scattering from a single particle $A$ or $B$ ``quark'' is
\begin{eqnarray}
W_{00}^{i}&=&\int d^3k \langle {\bf P}, i|\rho({\bf 0})|{\bf k}, i \rangle
\langle{\bf k}, i|\rho({\bf 0})|{\bf P}, i\rangle
(2\pi)^4\delta^4(P+q-k) \nonumber  \\
            &=&\langle {\bf P}, i|\rho_i({\bf 0})|{\bf P}+{\bf q},i\rangle
\langle{\bf P}+{\bf q},i|\rho_i({\bf 0})|{\bf P}, i\rangle \nonumber \\
&\times &(2\pi)^4 \delta(q^0-\frac{{\bf q}^2}{2m_i}) \nonumber \\
         &=&(2\pi)^{-2}\delta(q^0-\frac{{\bf q}^2}{2m_i}), i=A,B. \label{13}
\end{eqnarray}
The structure function for scattering from the ground state of the two-body
bound $AB$ system,  is
\begin{eqnarray}
W_{00}^{(AB)}&=&\sum_{\bf n}\int d^3k\langle
{\bf P}, {\bf 0}|\rho({\bf 0})|{\bf k}, {\bf n}\rangle
\langle {\bf k}, {\bf n}|\rho({\bf 0})|{\bf P}, {\bf 0}\rangle (2 \pi)^4
\delta^4(q+P-P_{{\bf k}, {\bf n}})  \nonumber \\
           &=&\sum_{\bf n}\langle {\bf P}, {\bf 0}
|\rho({\bf 0})|{\bf P}+{\bf q}, {\bf n}
\rangle \langle {\bf P}+{\bf q}, {\bf n}|\rho({\bf 0})|{\bf P}, {\bf 0}
\rangle \nonumber\\
&\times &(2 \pi)^4
\delta(q^0+E_{{\bf 0}}({\bf P})-E_{{\bf n}}({\bf P}+{\bf q})), \label{14}
\end{eqnarray}
where all states are $AB$ bound states.
If the difference $\epsilon_0 - \epsilon_{{\bf n}}$
could be neglected so that the $\delta$-function could be taken out of the sum,
completeness of the intermediate states would lead to
\begin{eqnarray}
W^{(AB)}_{00}&=&\sum_n \langle{\bf P},{\bf 0}|(\rho_A({\bf 0})+\rho_B({\bf 0}))
|{\bf P}+{\bf q}, {\bf n}\rangle \langle {\bf P}+{\bf q}, {\bf n}|
(\rho_A({\bf 0})+\rho_B({\bf 0}))|{\bf P},{\bf 0}\rangle \nonumber \\
& \times & (2\pi)^4
\delta(q^0-\frac{{\bf q}^2}{2(m_A+m_B)})  \label{15}
\end{eqnarray}
and each direct term would have the form
\[ \sum_n \langle {\bf P},{\bf 0}|\rho_i(0)|{\bf P}+{\bf q}, {\bf n}
\rangle \langle {\bf P}+{\bf q}, {\bf n}|\rho_i(0)|{\bf P},{\bf 0}\rangle
(2 \pi)^4 \delta(q^0-\frac{{\bf q}^2}{2(m_A+m_B)}) \]
\begin{equation}
=(2 \pi)^{-2}\delta(q^0-\frac{{\bf q}^2}{2(m_A+m_B)}),  \label{16}
\end{equation}
which is the structure function
for scattering from a particle of mass $m_A+m_B$.  This serves as a useful
check
on the calculation.
Taking into
account momentum conservation, the relevant matrix element is
\begin{eqnarray}
\langle{\bf P},{\bf 0}|\rho_{A}(0)|{\bf P}+{\bf q},{\bf n}\rangle&=&
(2 \pi)^{-3}\int F^{\ast}_{AB,{\bf 0}}({\bf y})
\exp(-\frac{i m_B{\bf q}\cdot {\bf y}}{m_A+m_B})F_{AB,{\bf n}}({\bf y})d^3y
\nonumber \\
        &=&(2 \pi)^{-3}\int F_{AB, 0}(y)
\exp(-\frac{i m_Bqy}{m_A+m_B})F_{AB, n}(y)dy \nonumber        \\
        &=&\frac{-i}{(2\pi)^3 \sqrt{n!}}\phi_A^{(n)},
\label{17}
\end{eqnarray}
\begin{equation}
\phi_A^{(n)}=(\frac{m_Bq^2}{2m_A(m_A+m_B)\hbar \omega})^{n/2}
exp(-\frac{m_B q^2}{4m_A(m_A+m_B)\hbar \omega}),    \label{18}
\end{equation}
${\bf q}$ was chosen in the $z$-direction, $q={\bf |q|}$, and $y$ is the
$z$-coordinate of ${\bf y}$.  With this choice, only the $z$-direction modes
of the higher oscillator states are excited.  A similar formula holds for
$\phi_B^{(n)}$.  Inserting Eq. (\ref{17}) in Eq. (\ref{14}),
\begin{equation}
W^{(AB)}_{00}=(2\pi)^{-2}\sum_n\frac{1}{n!}(\phi^{(n)}_A+\phi^{(n)}_B)^2
\delta(q^0-n\hbar \omega-\frac{{\bf q}^2}{2(m_A+m_B)}).  \label{19}
\end{equation}
Since, if the $n$ in the $\delta$-function could be neglected,
the sum on $n$ would cancel the exponential in each of the direct terms,
this verifies the check of Eq.(\ref{16}) above.
For this non-relativistic problem, I define the scaling variable to be
\begin{equation}
x={\bf q}^2/2(m_A+m_B)q^0.  \label{20}
\end{equation}
In terms of $x$, the structure function is
\begin{eqnarray}
W_{00}^{(AB)}&=&(2\pi)^{-2}\sum_n \frac{1}{n!}(\phi_A^{(n)}+\phi_B^{(n)})^2
\delta(\frac{q^2}{2(m_A+m_B)} \frac{(1-x)}{x} -n\hbar \omega)  \nonumber \\
&\approx & (2 \pi)^{-2}[exp(-{\cal Q}^2f_A(x))+exp(-{\cal Q}^2f_B(x))]^2
\nonumber \\
&\times & \sum_n \delta(({\cal Q}^2 \frac{(1-x)}{x} -n)\hbar \omega),
\label{21}
\end{eqnarray}
where ${\cal Q}^2\equiv q^2/2(m_A+m_B)\hbar \omega$ is dimensionless, and
\begin{equation}
2f_A(x)=\frac{m_A+m_B}{m_A} -\frac{1}{x} +\frac{1-x}{x} ln(\frac{m_A}{m_B}
\frac{1-x}{x}),  \label{24}
\end{equation}
and $f_B$ is the
same formula, except $A$ and $B$ are interchanged.  The
$\approx$ sign is because Stirling's approximation for $n!$ holds only for
large
$n$.
Each term of the sum comes from a different excitation of the $AB$ bound state.
These excitations play the role of resonances in hadronic physics.  The
different ``resonances'' contribute on disjoint lines of slope one in the
$q^0$--$q^2/2(m_A+m_B)$ plane.  In order to discuss the scaling limit, I
replace
the sum on $n$ by an integral over $n$.  Then
\begin{equation}
W_{AB}^{(00)}\approx \frac{1}{(2\pi)^2\hbar \omega}
(2 \pi)^{-2}[exp(-{\cal Q}^2f_A(x))+exp(-{\cal Q}^2f_B(x))]^2.
    \label{AB}
\end{equation}
The details of the averaging procedure do not matter.  For example, averaging
over $q^2$ at fixed $x$ gives the same result.
The functions $f_A$ and
$f_B$ are positive between $0 \leq x \leq 1$, except for quadratic zeroes at
\begin{equation}
x=m_A/(m_A+m_B)~{\rm and}~ x=m_B/(m_A+m_B),  \label{25}
\end{equation}
respectively; thus the structure
function vanishes for large $q$, except at these values of $x$ which are
the fractions of the momentum of the target bound state carried by the
respective ``quarks,'' as expected by the parton picture of deep inelastic
scattering.  Note also that, unless
$m_A=m_B$, $f_A+f_B$ has no zeroes; thus the interference term
vanishes in this limit,
which verifies the incoherence assumption of the parton picture\cite{4}.  The
rapid Gaussian decrease of $W_{00}^{(AB)}$ with
$q^2$ away from the values of $x$
at which scaling occurs reflects the rapid vanishing of the
``quark''-``quark'' potential at short distance mentioned in Sec. 1.

To evaluate $x$-moments of $W_{AB}^{(00)}$ in the large-$q^2$ limit,
approximate
$f_A$ by
\begin{equation}
f_A(x)\approx \frac{(m_A+m_B)^4}{4m_A^3m_B}(x-\frac{m_A}{m_A+m_B}).
\end{equation}
Then
\begin{equation}
\int_0^1 W_{00}^{(AB)} dx\approx \frac{2\sqrt{\pi m_Am_B}}{(2 \pi)^2
\hbar \omega (m_A+m_B){\cal Q}};
\end{equation}
thus moments of $q^0W_{00}^{(AB)}$ remain finite and non-vanishing in the
scaling limit.  This result holds for either order of doing $\int dx$ and
replacing $\sum_n$ by $\int dn$.  However, if the scaling limit, $q^2
\rightarrow \infty$, is (improperly) taken before calculating the $x$-moments,
then, since $W_{00}^{(AB)}$ is bounded and vanishes in this limit, except at
the two isolated
points given by Eq.(\ref{25}), the $x$-moments would appear to
vanish\cite{wallace}.

Scaling in deep inelastic scattering from the deuteron considered as a bound
state of a proton and a neutron, with emphasis on the effect of Fermi
motion, was discussed in\cite{west}.

The calculation of $W_{00}^{(AB)}$
in Eq.(\ref{19}) corresponds to Figure 1; the
parton model limit corresponds to Figure 2.  Note that Figure 1 is a two-loop
graph while Figure 2 is a one loop graph; thus in the scaling limit the graph
with
two independent momentum integrations reduces to a single such integration.
Figure 3 shows the fixed-$x$ and
fixed-resonance mass lines in the $q^0$--$q^2/2(m_A+m_B)$ plane.
\vglue.2in

\begin{center}
{\bf 4. SUMMARY AND OUTLOOK}
\end{center}

In a simple model, I verified that the deep inelastic limit of the
structure function approaches the limit of incoherent elastic scattering off
its constituents as though the constituents were free.  This shows the way in
which the scaling limit can coexist with confinement.  In the present two-body
model, each constituent ``quark'' carries a fixed part of the total momentum.
It would be interesting to study a three-body bound state in which the
``quarks'' can carry variable fractions of the total momentum, as well as a
model in which there are a variable number of constituents.
The three ``quark'' masses, the Lagrangian mass which occurs in the kinetic
term in the Hamiltonian
(or the corresponding Lagrangian), Eq.(\ref{1}), the
constituent mass which occurs in the (analog of the) Schr\"odinger equation,
Eq.(\ref{4}), and the current mass which occurs in the parton model limit,
Eq.(\ref{20}),
are all the same in this simple model.  These masses will differ in more
realistic models.

\begin{center}
{\bf ACKNOWLEDGEMENT}
\end{center}

It is a pleasure to thank Boris Ioffe for suggesting this calculation and
for very helpful discussions about the work, and to thank Manoj Banerjee, Tom
Cohen and Steve Wallace for reading the manuscript and making useful
suggestions.

\begin{center}
{\bf APPENDIX: REVIEW OF THE CONFINED IN FIELD VERSION\\
 OF THE N QUANTUM APPROXIMATION.}
\end{center}

The basic idea of the N quantum approximation is that the complete and
irreducible set of Heisenberg fields which appear in the Lagrangian of a
theory can be expanded,
following the seminal paper of Rudolf Haag \cite{haag},
in (normal-ordered, if one chooses) products of the
complete and irreducible set of asymptotic fields for the stable particles
of the theory, both the particles which correspond to the Heisenberg fields
and the particles corresponding to any bound states which are present.  The
c-number coefficients of the Haag expansion (I call these the Haag amplitudes)
are retarded
(if in fields are chosen)
or advanced (if out fields are chosen) amplitudes with the legs corresponding
to the asymptotic fields on shell and the single leg corresponding to the
Heisenberg field off shell.  Thus only one leg in any Haag
amplitude is off shell.
This formalism is as close to being on shell as can be achieved in a field
theory.  The Haag amplitudes are closely related to scattering and bound state
amplitudes.  For theories with confined particles, I assume that a state with
only one confined particle is allowed, but that scattering states
of more than one confined particle are prohibited.
In order to study the modifications necessary in the Haag expansion of the
Heisenberg fields in normal-ordered in (or out) fields for the case of theories
with confinement, I studied a model \cite{con}
in which non-relativistic particles, called
``quarks,'' interact with a harmonic potential.  Solution of the model required
that the in (or out) fields be replaced by ``confined in (or out) fields''
which differ from the usual asymptotic fields by having vacuum projectors
$\Lambda_0$ to
the left of the annihilation parts of the asymptotic fields and to the right
of the creation parts of the asymptotic fields.  The insertion of vacuum
(or in quantum chromodynamics (QCD), color-singlet) projectors enforces the
prohibition of multiparticle scattering states.  For the model of \cite{con}
with
confinement by a harmonic potential there are no analogs of the unconfined
mesons and baryons of QCD,  so the in and out fields are identical.
For a realistic model with color and
antiquarks, the vacuum projectors should be replaced by projectors onto the
color-singlet states.

Although I use the
confined in field formalism, as emphasized in Sec. 1, the results do not depend
on the use of this
formalism.

The relevant terms in the Haag expansion, with confined in fields, are
\begin{equation}
\psi_A({\bf x},t)=\Lambda_0 \psi_{A,in}({\bf x},t)+\sum_{{\bf n}}
\int F_{AB,{\bf n}}({\bf x}-{\bf y})\psi_{B,in}^{\dagger}({\bf y},t)
\Lambda_0 B_{{\bf n},in}({\bf R},t)d^3y,                          \label{3}
\end{equation}
where ${\bf R}=(m_A{\bf x}+m_B{\bf y})/(m_A+m_B)$, $\psi_{A,in}$ and
$\psi_{B,in}$ are Fermi in fields and obey Fermi equal-time
anticommutation relations (for this non-relativistic model, either Fermi or
Bose statistics could be chosen),
and $B_{{\bf n},in}$ is the Bose
in field for the two-body bound state in oscillator level ${\bf n}$ and obeys
Bose equal-time commutation relations.
There is a similar equation for the field $\psi_B$ with $A$ and $B$
interchanged and $F_{BA,{\bf n}}({\bf x})=-F_{AB,{\bf n}}(-{\bf x})$.
The form of the expansion for the terms with bound states is dictated by the
requirements of translation and
Galilean invariance.  Focusing attention on the
two-body bound states, the Haag amplitudes $F_{AB,{\bf n}}$,
${\bf n}=(n_1,n_2,n_3)$, satisfy the
Schr\"odinger equation, and thus
are the Schr\"odinger amplitudes for the oscillator bound states with energy
$\epsilon_{{\bf n}}=\sum_i (n_i+3/2)\hbar \omega$,
$\hbar \omega = \sqrt{k/2 \mu_{AB}}$,
at rest.

The expansions of the fields in annihilation operators are
\begin{equation}
\psi_i({\bf x},t)=(2 \pi)^{-3/2}\int d^3p dE a_i({\bf p},E)
exp(-iEt+i{\bf p \cdot x}),~ i=A,B,     \label{8}
\end{equation}
\begin{equation}
\psi_{i,in}({\bf x},t)=(2 \pi)^{-3/2}\int d^3p a_{i,in}({\bf p})
exp(-i{\bf p}^2t/2m_i+i{\bf p} \cdot {\bf x}),~ i=A,B,      \label{9}
\end{equation}
\begin{equation}
B_{{\bf n},in}({\bf x},t)=(2 \pi)^{-3/2}\int d^3p \Lambda_0 b_{{\bf n},in}
({\bf p}) exp(-iE_{{\bf n}}({\bf p})t+i{\bf p} \cdot {\bf x}), \label{10}
\end{equation}
$E_{{\bf n}}({\bf p})={\bf p}^2/2(m_A+m_B)+\epsilon_{\bf n}$ for the
two-body bound states.  The equal-time anticommutation relations for the in
fields, $[\psi_i({\bf x},t),\psi^{\dagger}_j({\bf y},t)]_+=
\delta_{ij}\delta({\bf x}-{\bf y})$,
lead to anticommutation relations for
the annihilation and creation operators,\\
$[a_{i,in}({\bf p}), a^{\dagger}_{j,in}({\bf q})]_+=
\delta_{ij}\delta({\bf p}-{\bf q})$.
The charge density $j_0(x)\equiv \rho(x)$ is
$\rho(x)=\rho_A(x)+\rho_B(x)$,
\begin{eqnarray}
\rho_A(x)&=&\psi_A^{\dagger}(x)\psi_A(x)   \nonumber \\
         &=&\psi_{A,in}^{\dagger}(x)\Lambda_0\psi_{A,in}(x)   \nonumber \\
        &+ &\sum_{{\bf n},{\bf n^{\prime}}}
\int F_{AB,{\bf n}}^{\ast}({\bf x}-{\bf y})
B_{{\bf n},in}^{\dagger}({\bf R},t)\Lambda_0 B_{{\bf n^{\prime}},in}({\bf R},t)
F_{AB,{\bf n^{\prime}}}({\bf x}-{\bf y})d^3y,          \label{11}
\end{eqnarray}
where I have kept only the relevant terms in $\rho_A$.
\begin{center}
{\bf FIGURE CAPTIONS}
\end{center}

Fig. 1:  Graph for $W_{00}$ in which light lines are on shell, heavy lines are
off shell and the sum stands for the integrals over ${\bf k}$ and
${\bf k}^{\prime}$ and a sum over harmonic levels $n$.  The line labeled by
the momenta $P$ and $P+q$ are two-body bound states; the other solid lines are
single-particle states.  The wiggly lines are currents.

Fig. 2:  Graph for the parton model limit of $W_{00}$.  Here there is just one
integration, over ${\bf k}$.  The line labeled by $P$ is the two-body ground
state; the other solid lines are single-particle states.  The wiggly lines are
again currents.

Fig. 3:  The allowed kinematic domain for the structure function is between the
$45^{\circ}$, line $x=1$, and the $q^0$ axis, $x=0$.  Generic fixed $x$ and
generic fixed resonance mass $M_r$ are indicated.

\end{document}